\documentclass{article}
\usepackage{mrs2005,epsfig}
\begin{document} 
\title{INTERNAL KINEMATICS AND STELLAR POPULATIONS OF EARLY-TYPE GALAXIES IN THE FORNAX CLUSTER}

\author{Bettina Gerken$^{1,2}$, Harald Kuntschner$^1$, Roger L. Davies$^2$}

\affil{$^1$European Southern Observatory, Karl-Schwarzschild-Str. 2, 85748 Garching near Munich, Germany \\ 
       $^2$University of Oxford, Department of Astrophysics, Denys Wilkinson Building, Keble Road, Oxford OX1 3RH, United Kingdom} 
  
\begin{abstract} 
We present a study of the internal kinematics and stellar populations of early-type galaxies in the Fornax cluster based on integral-field spectroscopic observations with Gemini South GMOS and VLT-VIMOS. Seven galaxies in a luminosity range of -21.3$\le$M$_B$$\le$-17.7 have been observed with these integral field units (IFU). As first results, velocity and line strength maps are presented for NGC\,1427 and NGC\,1419.
\end{abstract} 
 
\section{Introduction} 
 Elliptical galaxies and bulges divide into two groups with fundamental differences in their morphological, dynamic and core profile properties \cite{KB96,FTA97}. Luminous systems rotate slowly, have boxy isophotes and cores, whereas faint early-type galaxies show rapid rotation, discy isophotes and a cuspy central luminosity profile. These differences can be interpreted as differences in the formation of both classes which may be reflected in their stellar populations. Studies of stellar populations have revealed signs of secondary star formation in some lenticular galaxies \cite{KD98}. Many early-type galaxies show kinematic peculiarities such as nuclear stellar and gaseous discs, kinematically decoupled components (KDC) or minor axis rotation \cite{ECP04}. A combination of 2-dimensional kinematics and stellar population analysis provides information about the formation history of early-type galaxies.

\section{The Fornax IFU survey}
The Fornax cluster galaxies NGC\,1427, NGC\,1380, NGC\,1381 were observed with the GMOS IFU, and a second sample of NGC\,1339, NGC\,1375, NGC\,1419 and ESO\,358-G25 and additionally NGC\,3379 were observed with the VIMOS IFU. The VIMOS IFU has a large field of view of $27''\times27''$ at a spatial resolution of $0.67''$, whereas GMOS has a smaller field of view of $5''\times7''$ but offers a spatial resolution of $0.14''$. This approach allows a study of the larger kinematic scales and to detect possible line strength gradients in the VIMOS sample, while on the other hand the very centres of the galaxies observed with GMOS can be studied in great detail. Relations between the internal kinematics and stellar populations are indicative of the star formation history of the host galaxy. The wavelength range of the combined sample, approximately 4200\,\AA\ to 5500\,\AA, includes H$\beta$ and higher order Balmer lines, allowing an age measurement largely insensitive to metallicity effects. Several Fe indices and Mg line strengths are measured to determine the metallicity and [Mg/Fe] ratio of the stellar populations.  

%  
% Figure 1 
% 
\begin{figure}  
%\vspace*{1.25cm}  
\begin{center}
\epsfig{figure=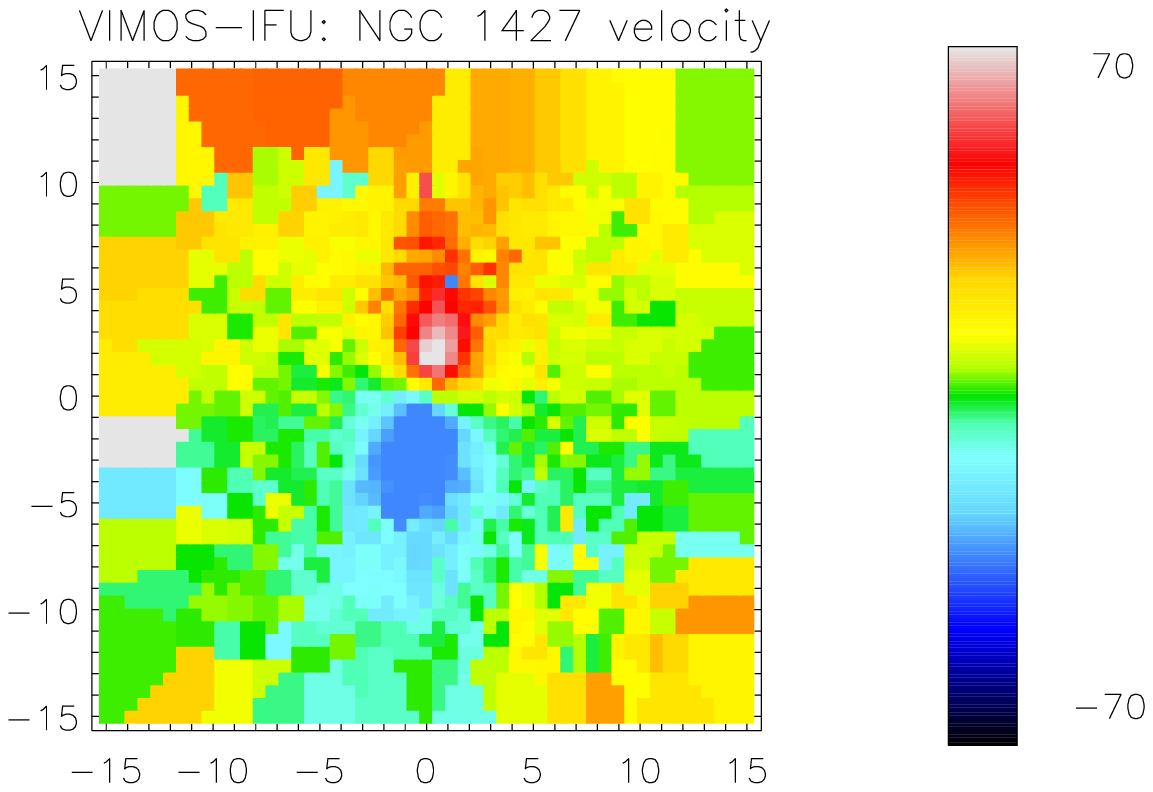,width=6.5cm}
\epsfig{figure=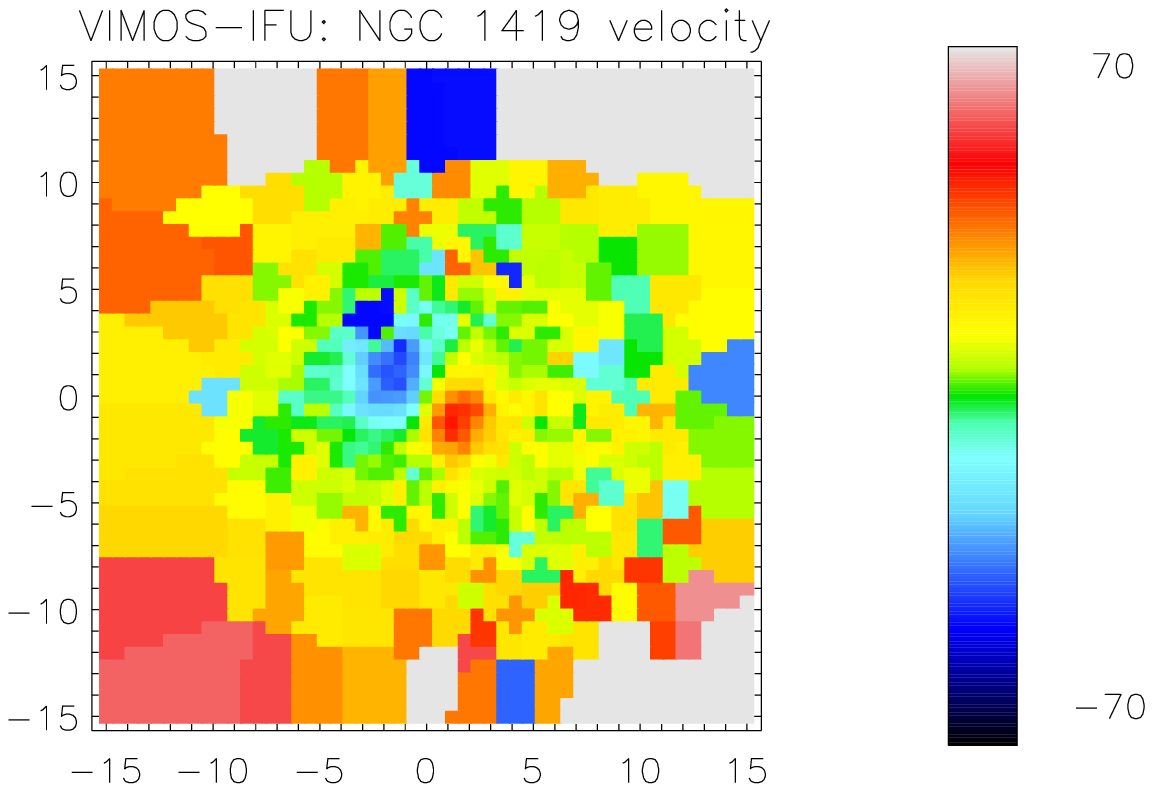,width=6.5cm} 
\end{center}
%\vspace*{0.25cm}  
\caption{  Velocity maps of NGC\,1419 (left) and NGC\,1427 (right). The velocity ranges in km/s and colour scales are shown to the right of the maps. Both maps are binned to a minimum S/N of 50. 
} 
\label{fig:vmaps}
\end{figure} 

\begin{figure}  
%\vspace*{1.25cm}  
\begin{center}
\epsfig{figure=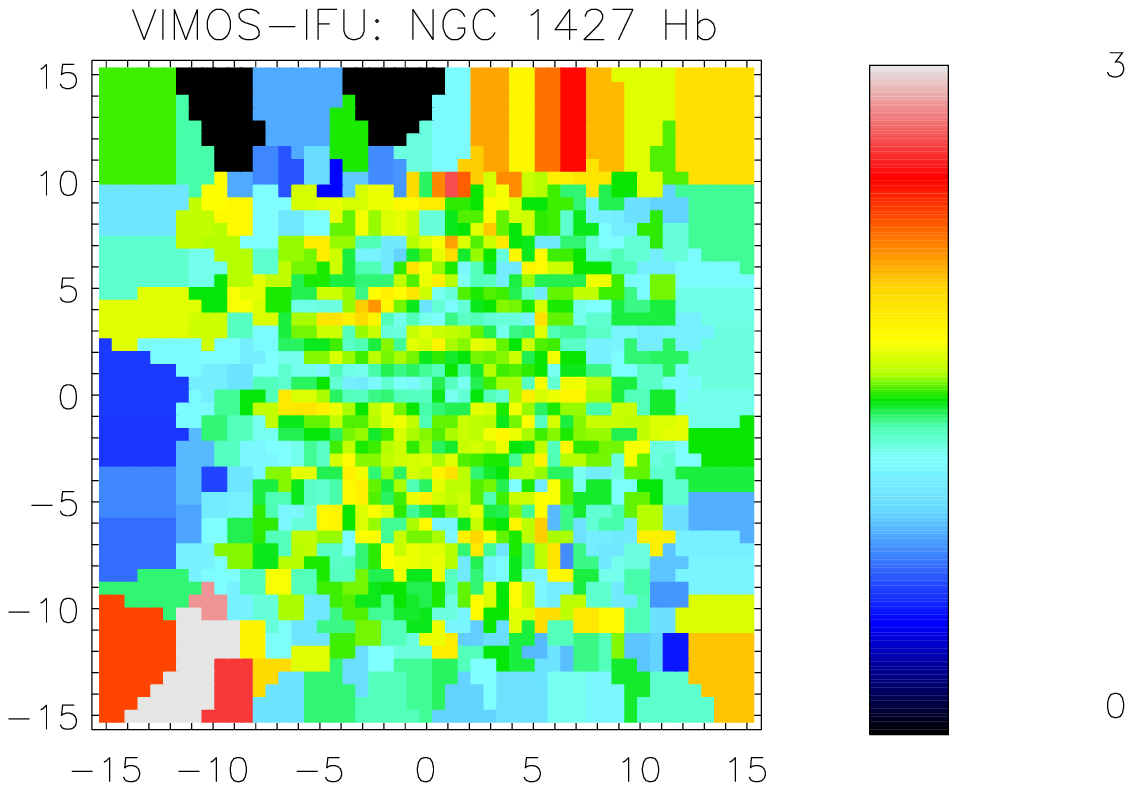,width=6.5cm}
\epsfig{figure=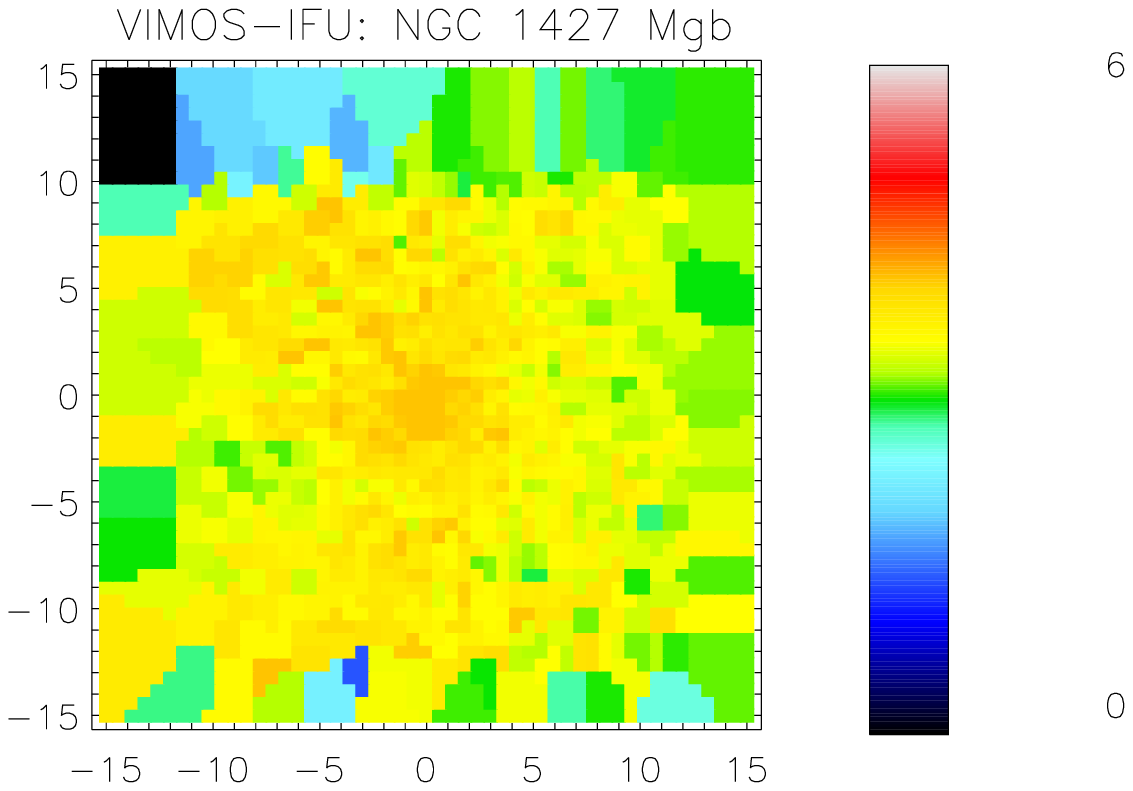,width=6.5cm}
\epsfig{figure=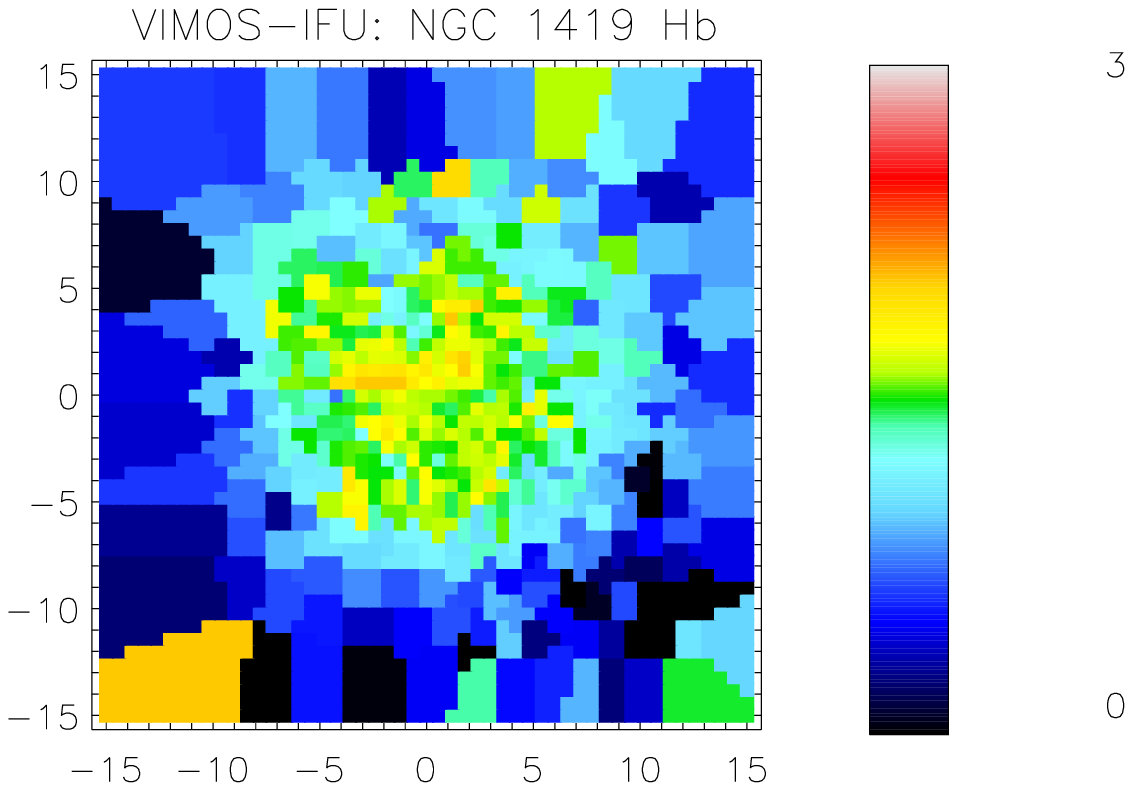,width=6.5cm}
\epsfig{figure=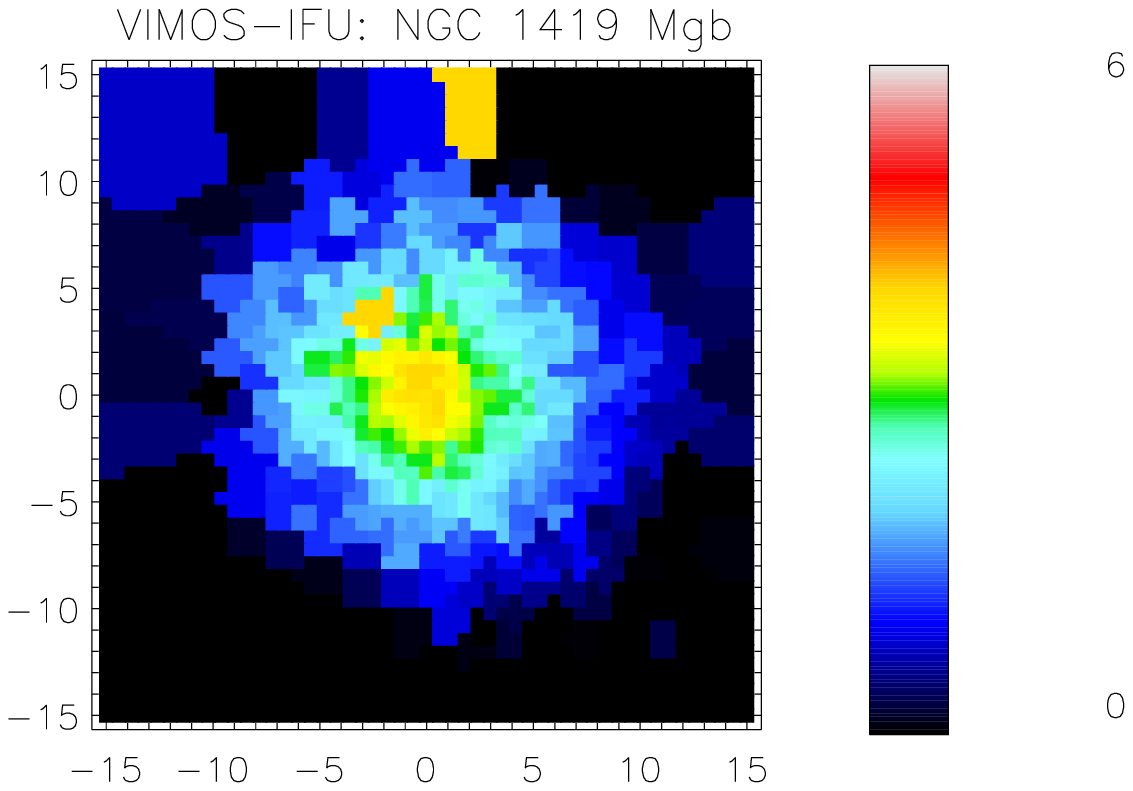,width=6.5cm}
\end{center}
%\vspace*{0.25cm}  
\caption{ Line strength maps of NGC\,1427 (top panels) and NGC\,1419 (bottom panels). The indices shown are H$\beta$ (left) and Mg$b$ (right). The index ranges in \AA\ and colour scales are given to the right of the maps. All maps are binned to a minimum S/N of 50.
} 
\label{fig:linemaps}
\end{figure} 

%\section{} 

\section{Results} 

Fig.~\ref{fig:vmaps} shows the stellar velocity maps of NGC\,1427 (left panel) and NGC\,1419 (right panel). The known decoupled core of NGC\,1427 is clearly visible. For the first time, a kinematically decoupled core is also detected in NGC\,1419. Line strength maps for the same galaxies are shown in Fig.~\ref{fig:linemaps}. The top panels show the H$\beta$ (left) and Mg$b$ (right) distribution in NGC\,1427, while the bottom panels show the same for NGC\,1419. In the H$\beta$ distribution, there is an increase towards the centre visible in NGC\,1419, whereas the distribution is almost flat in NGC\,1427. Both galaxies show an Mg$b$ gradient with increasing line strengths towards the centre, which is more pronounced in NGC\,1419. A comparison between the velocity and line strength maps implies that the stellar populations in the decoupled components do not differ significantly from the main body of the galaxies.

\acknowledgements{ We would like to thank Bryan Miller and the Gemini South staff for efficient support during the observations and with the data reduction.
}

\vfill 
\end{document}